# HOMO band structure and anisotropic effective hole mass in thin crystalline Pentacene films

Richard C. Hatch[1], David L. Huber[2] and Hartmut Höchst[1,*]

[1]Synchrotron Radiation Center, University of Wisconsin-Madison, 3731 Schneider Dr. Stoughton, WI 53589

[2]Department of Physics, University of Wisconsin-Madison, 1150 University Avenue, Madison, WI 53706

The band dispersion of the two highest occupied molecular orbital (HOMO)-derived bands in thin crystalline Pentacene films grown on Bi(001) was determined by photoemission spectroscopy. Compared to first-principles calculations our data show a significantly smaller band width and a much larger band separation indicating that the molecular interactions are weaker than predicted by theory—a direct contradiction to previous reports by Kakuta *et al*. [Phys. Rev. Lett. **98**, 247601 (2007)]. The effective hole mass $m^*$ at $\overline{M}$ is found to be anisotropic and larger than theoretically predicted. Comparison of $m^*$ to field effect mobility measurements shows that the band structure has a strong influence on the mobility even at room temperature.

Recently much research has been performed on organic semiconductors (OSCs) because of their promising device properties.[1,2] Pentacene (Pn), with mobilities rivaling those of amorphous Si,[3] is a prototypical OSC. Despite efforts of theoreticians and experimenters the transport mechanism in Pn and other OSCs is not fully understood. OSCs are reported to have band like transport at low temperatures.[4-7] As the temperature increases, however, the charge carrier transport mainly occurs by variable-range hopping.[8,9] Furthermore, it is predicted that at room temperature (RT) the softness of these materials allows thermal molecular motion to destroy the translational symmetry of the Hamiltonian and a localization of charge carriers results.[10] There are works, however, that suggest that the band structure should have an influence on transport at RT even though the charge carriers are scattered by lattice vibrations.[11-13] Recent photoemission experiments by Kakuta *et al.*[11] concluded that the intermolecular interaction in Pn is much stronger and leads to a larger band dispersion than predicted by theory. Our experiments on Pn thin films are in contradiction to the data and conclusions of Kakuta *et al.*[11]

We prepared Pn films on Bi(001) and studied the band dispersions with angle resolved photoemission spectroscopy (ARPES). The Bi(001) substrates were grown by depositing 100 Å of Bi, onto Si(111)–7×7 at RT. The Bi films were annealed *in situ* at 460 K for 8 hours to further improve the crystalline quality[14] as evidenced by the increased sharpness of reflection high energy electron diffraction (RHEED) patterns and in the increase of the very pronounced Bi surface-states emission features.[15] Pn films were grown out of an effusion cell onto Bi(001) at 345 K. The geometry of the Pn source and the Bi(001) substrate was such that the Pn flux was in the Bi[$11\bar{2}$] direction at an incident angle of ~10°. The grazing incidence growth geometry was reported to enhance the domain size and stimulate a certain growth direction relative to the substrate by step-flow growth.[16]

Pn grows on Bi(001) with the *ab*-plane parallel to the surface and the *a*-axis aligned with the Bi[$1\bar{1}0$] direction. The *b*-axis is free to orient itself 4° to the left or right of the Bi[$11\bar{2}$] direction resulting in an angle between the *a*- and *b*-axes of $\gamma$ = 86° or 94° for the chiral twin.[17] The second molecule in the unit cell is located at (*a*+*b*)/2, and has a different orientation giving rise to the "staircase" pattern explained in Ref. 18. Due to the symmetry of the Bi(001) surface and the Pn surface Brillouin zone (SBZ) there are six effective orientations for Pn domains: three different rotations and twins for each rotation. By using grazing incidence deposition larger domains were grown with the *a*-axis aligned with the Bi[$1\bar{1}0$] and the *b*-axis 4° to the right of the Bi[$11\bar{2}$]. From RHEED we verified the lattice parameters to be $a$ = 6.0 ± 0.2 Å, $b$ = 7.9 ± 0.2 Å and $\gamma$ = 86 ± 2°, consistent with literature.[17] Other Pn domains were also present, however, as we will discuss later these domains were smaller in quantity and/or size and their contribution did not drastically affect the analysis of the photoemission data.

The nominal thickness of the Pn films as determined with a quartz microbalance was ~400 Å. In all cases the Pn films were thick enough to completely suppress all Bi-derived photoemission features. ARPES experiments were performed on *in situ* grown films at the University of Wisconsin Synchrotron Radiation Center (SRC). The combined photon and electron energy resolution was ΔE ~40 meV. For 15 eV photons the electron momentum resolution was $\Delta k_{\parallel} < 0.03$ Å$^{-1}$. All samples were measured at 75 K unless otherwise noted.

Since the electronic properties of interest in OSC's are closely related to the upper most valence bands, we focused our study on the dispersion of the two highest occupied molecular orbital (HOMO)-derived bands, which will be referred to as the HOMO1 and HOMO2 bands for the lower and higher binding energy bands respectively. Figure 1 shows the SBZ of Pn and locations (red dots) at which ARPES data were collected.

Figure 2(a) shows photoemission spectra measured with $hv$ = 15 eV near the top of the valence band. Data taken along Pa$_1$ clearly show the evolution of a double peak structure. To visualize the dispersive nature of the valence band features we took the second derivative of the data whose negative part is shown in Fig. 2(b). Red diamonds mark the intensity minima of the derivative spectra. For a quantitative analysis we fit a set of Gaussian peaks to the individual spectra. The energy positions of the most prominent Gaussians and their respective parallel electron momentum k$_{\parallel}$ are shown in Fig. 2(c). Additional data in other crystallographic directions are given in the online supporting material.[19] The valence band maximum of the HOMO emission is at $\overline{M}$ and has a binding energy of $E_B$ = -1.15 eV. Similar values are reported for Pn grown on a variety of metallic substrates.[20] Changing the photon energy alters the relative intensities of the HOMO valence band emission but does not change the k$_{\parallel}$ band dispersion, indicating that the band structure of our films is primarily two dimensional with no noticeable dispersion along the perpendicular component of **k**.

The experimental band dispersions as shown in Fig. 2(c) differ strongly from similar experiments[11, 12] involving Pn films. Pn grown on highly oriented graphite revealed an unstructured broad photoemission peak whose angular dependence could be analyzed with a single Gaussian function[12], while ARPES from a 1 monolayer Pn film by Kakuta *et al.* shows the emergence of a second feature for off normal emission.[11] Their photoemission data with a virtually non-dispersive HOMO1 band does not support our findings. In a separate study we verified that the observed discrepancies are not exclusively related to substrate interactions that might be more noticeable in the much thinner Pn film of Ref. 11.

To further analyze our data we chose a tight-binding model as suggested in[11, 12, 18, 21] to describe the dispersion of the two HOMO derived bands. The band dispersions are given by

$$E_{\pm}(\mathbf{k}) = E_0 + T_1(\mathbf{k}) \pm \left[T_2^2(\mathbf{k}) + (\Delta E/2)^2\right]^{1/2} \qquad (1)$$

where **k** is the wave vector and $\Delta E = E_1 - E_2$ accounts for the energy difference between the two inequivalent sites. $E_0 = (E_1+E_2)/2$ is their average energy. Molecular interactions are represented by $T_{1,2}(\mathbf{k})$ which incorporates exchanges between molecules in the same and in different sub-lattices respectively. $T_{1,2}(\mathbf{k})$ contains transfer integrals $t_i^{1,2}$ given by

$$T_{1,2}(\mathbf{k}) = 2\sum_i t_i^{1,2} \cos(\mathbf{k} \cdot \mathbf{r}_i^{1,2}) \qquad (2)$$

where $\mathbf{r}_i^{1,2}$ are the vectors defining the inter-molecular separations.

The tight binding fit parameters are given in Table I. As one can see from Table I the HOMO-derived band dispersions are mostly affected by intermolecular interactions corresponding to the largest transfer integrals which are the nearest neighbor interactions along the vectors (*a-b*)/2, (*a+b*)/2 and *a*.

Figure 3(a) compares our photoemission based tight binding band dispersion with previously mentioned first-principle band structure calculations.[18, 21] The experimental and theoretical critical point energies at the $\overline{\Gamma}$ point are similar, but the overall band dispersions and energies differ, most notably in the $\overline{M}$ and $\overline{Y}$ regions. The first-principles DFT calculation[18] shows roughly twice the band widths in each of the two HOMO bands and about one third the separation between the two bands near $\overline{X}$, $\overline{Y}$, and along $\overline{\Gamma}$-$\overline{M}$ suggesting that DFT overestimates the strength of the intermolecular interactions. The calculation in Ref. 21 of the bulk film Pn polymorph shows a 25% larger band width in HOMO1 but a 25% smaller band width in HOMO2. Because a band width narrowing with increasing temperature has been predicted in organic molecular crystals[22] a few spectra were taken at RT which verified that these large discrepancies are not temperature related.

The fact that first principles calculations and the experimental band structure differ significantly even though two of the leading transfer integrals are in reasonable agreement emphasizes that this overlap is not purely a function of intermolecular distances, but can be strongly affected by the different stacking of Pn molecules in various polymorphs.[21]

Figure 3(b) compares this work (red line) with results from Ref. 11 (green dots). The contrast between the two works is apparent, most notably along the $\overline{\Gamma}$ - $\overline{M}$ and $\overline{\Gamma}$ - $\overline{X}$ directions of HOMO1 as well as along the $\overline{\Gamma}$ - $\overline{Y}$ direction of HOMO2. There are two most likely sources of these discrepancies. First, Kakuta *et al*. performed experiments on a single monolayer of Pn (~15 Å) which may interact with the Bi substrate and Bi-derived photoemission features may hinder the accompanying analysis. This is in contrast to experiments performed in this work where Pn-film thicknesses were hundreds of Ångstroms thick and much more representative of bulk Pn. Second, our Pn films were grown on a 345 K Bi substrate which we found resulted in a more crystalline Pn film than when grown at RT as in Ref. 11.

As mentioned earlier, Pn films generally contain domains of various size and orientation. Although we can not completely rule out multi-domain contributions, there are several indications that our data are from mostly a single domain. The presence of only two features in the second derivatives (Fig. 2(b)) is the strongest indication against significant contributions from the other two rotated domains. The influence of a chiral twin is a little more difficult to rule out, but there is evidence suggesting that the $\gamma = 94°$ twin is absent. First is the minimum in HOMO1 along Pa$_1$ occurring at k$_{||}$ ~0.2 Å$^{-1}$ that is absent along Pa$_2$ (see Fig. 2(c)). This minimum which is predicted by theory[18, 21] is seen in the peak positions, as well as in the tight-binding fit to the data. If both chiral twins were present to an equal degree, the spectra along $Pa_1$ and $Pa_2$ would be virtually identical, which they are not. Second, we found fitting Eqs. (1-2) to the experimental data with $\gamma = 86°$ resulted in a noticeably better least-squares fit than choosing the chiral twin with $\gamma = 94°$.

As previously mentioned, the carrier mobilities in OSCs are of great importance for any device application. The hole mobility is given by $\boldsymbol{\mu}_h = e\tau / \boldsymbol{m}^*$ where $\boldsymbol{m}^*$ is the effective hole mass tensor and $\tau$ the isotropic total scattering time. Using the tight binding fit to the HOMO1 band and the relation $-\partial^2 E(\mathbf{k})/\partial \mathbf{k}^2 = \hbar^2/\boldsymbol{m}^*$, we calculated $1/\boldsymbol{m}^*$. The result for the effective hole mass at the $\overline{M}$-point, which is the valence band maximum and as such should have the largest affect on $\boldsymbol{\mu}_h$, is shown in Fig. 4. Comparison with theory shows that the shape, magnitude and orientation of $1/\boldsymbol{m}^*$ are not dramatically different than those predicted by theory for Pn single crystals.[18] However, in addition to a roughly 10° misalignment the data also show a more pronounced anisotropy of $1/\boldsymbol{m}^*$.

Also shown is the effective mass that we extracted from mobility data reported in Ref. 23. The $\mathbf{k}$ dependence of the $\boldsymbol{m}^*$ tensor suggests that the band structure plays a major role in Pn even at RT. Since the mobility measurement[23] was not referenced with respect to the Pn lattice vectors we rotated and scaled the data to best match our experimental curve of $1/\boldsymbol{m}^*$. The resulting scaling factor corresponds to an isotropic scattering time, $\tau$ ~3 fs at 300 K.

In conclusion, Pn thin films of high structural quality were grown on Bi(001) substrates with domains large enough for ARPES to reveal the HOMO-derived Pn in-plane band dispersions. The experimentally determined band structure, when compared to theory,[18, 21] shows that current theoretical models overestimate the interactions between molecules in Pn crystals which is in direct contradiction to previous experiments performed by Kakuta *et al.*[11] on a single monolayer film. A tight binding model was fit to the ARPES band structure from which we extracted the HOMO-derived Pn in-plane band dispersions. The effective mass *m*\* extracted at $\overline{M}$ is smaller and more directionally dependent than predicted by theory.[18] A comparison of 1/*m*\* with mobility measurements[23] emphasizes that the band structure has a strong role in transport properties in Pn.

We acknowledge the help of undergraduate researchers, Justin Frederick and Emily Kopp. The SRC, University of Wisconsin-Madison, is supported by the NSF under Award No. DMR-0537588.

*Corresponding author: hhochst@wisc.edu

TABLE I. Parameters obtained from a tight-binding model [Eqs. (1-2)] fit to the experimental Pn band structure. All values are in meV. Also shown are the fit values to the energy bands of a DFT calculation for a Pn single crystal in Ref. 18, and the parameters reported for two Pn polymorphs whose band structure was calculated in Ref. 21. The subscripts on the transfer integrals define the intermolecular separation.

| | This Work | Single Crystal[a] | Single Crystal[b] | Bulk Film[b] |
|---|---|---|---|---|
| $E_0$ | -1327 ± 1 | | | |
| $\Delta E$ | 134 ± 3 | 33 | 38 | 24 |
| $t_{(a+b)/2}$ | -47 ± 1 | -56 | -43 | -63 |
| $t_{(a-b)/2}$ | 56 ± 1 | 87 | 79 | 24 |
| $t_a$ | 18.3 ± 0.5 | 32 | 24 | 1 |
| $t_b$ | 1.6 ± 0.5 | -6 | -1 | -3 |
| $t_{(a+b)}$ | -2.6 ± 0.5 | -1 | 2 | 0 |
| $t_{(a-b)}$ | 2.9 ± 0.5 | 2 | 0 | 3 |
| $t_{2a}$ | 0 ± 0.5 | -1 | -1 | -1 |
| $t_{2b}$ | 1.1 ± 0.5 | -1 | 0 | 0 |

[a]Ref. 18, [b]Ref. 21

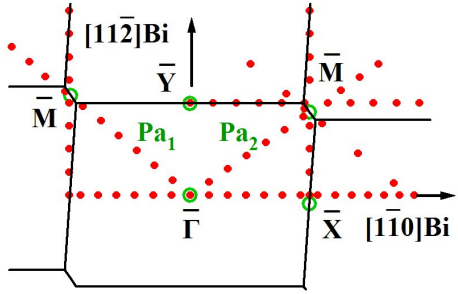

FIG. 1 (color online). Orientation of the Pn SBZ (black) with respect to the underlying Bi(001) substrate. Symmetry points are shown with green open circles. $Pa_1$ and $Pa_2$ are paths along the respective ($a^*-b^*$) and ($a^*+b^*$) directions, where $a^*$ and $b^*$ are reciprocal lattice vectors. Red dots mark the $k_{x,y}$ location where ARPES data were taken.

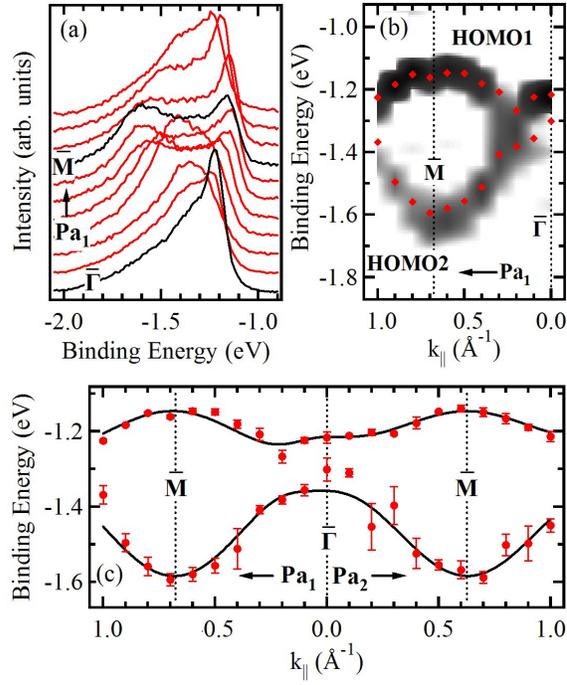

FIG. 2 (color online). (a) ARPES along $Pa_1$ from $k_\parallel = 0$ to 1 Å$^{-1}$ in steps of ~0.1 Å$^{-1}$ displays dispersing features corresponding to the HOMO-derived Pn bands. (b) Negative part of second derivatives of spectra shown in (a) to visualize the two dispersing branches of the HOMO-derived bands. (c) Tight-binding fit (black line) to the ARPES peak positions (red dots). See text for details.

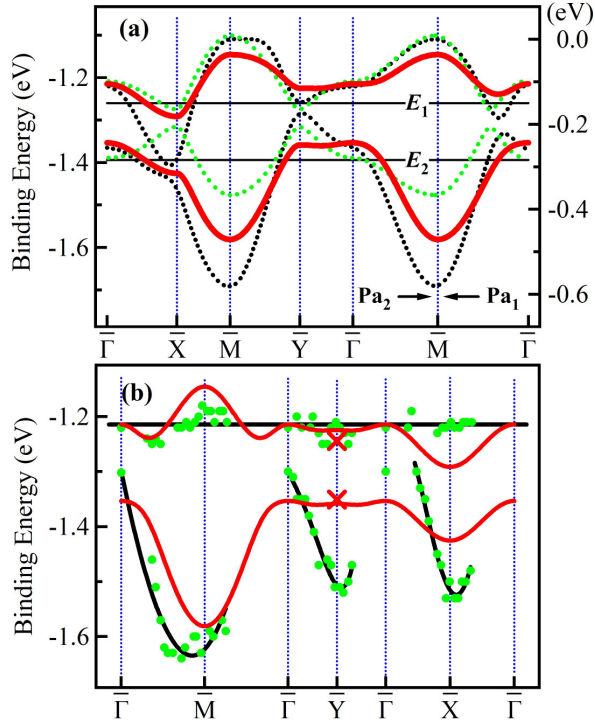

FIG. 3 (color online). (a) Comparison of tight-binding fit to Pn-ARPES band structure (red (gray) lines) with theoretical band structures for single crystal[18] and bulk film phase[21] Pn-polymorphs (black and green (gray) dotted lines respectively). The horizontal lines $E_1$ and $E_2$ are the energies of the two inequivalent sites. (b) Data from Ref. 11 (green dots) with their general trends[19] (black lines) compared to the Pn-ARPES band structure derived in this work (red (gray) lines). The band energies at $\bar{Y}$ determined from our ARPES are shown with red crosses, and within error bars are identical to those determined at $\bar{\Gamma}$. This suggests that a detailed mapping of the band structure along this direction is unnecessary.

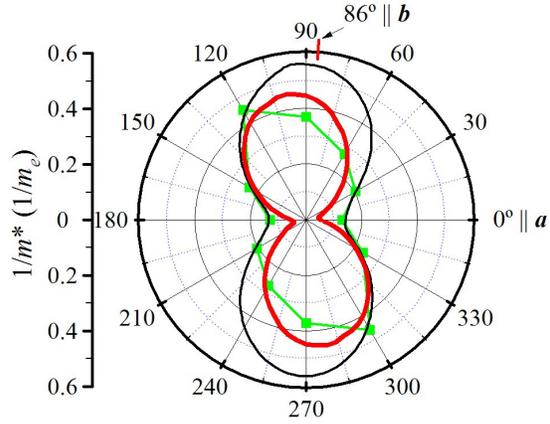

FIG. 4 (color online). Inverse effective hole mass 1/$m^*$ of Pn. Red (gray) line: extracted from our experimental data. Black line: single crystal theory.[18] Green squares: derived from field effect mobility measurement[23] at ~300 K assuming isotropic scattering.

# Appendix

**Supplementary information for manuscript "HOMO band structure and anisotropic effective hole mass in thin crystalline Pentacene films" by Hatch *et al*.**

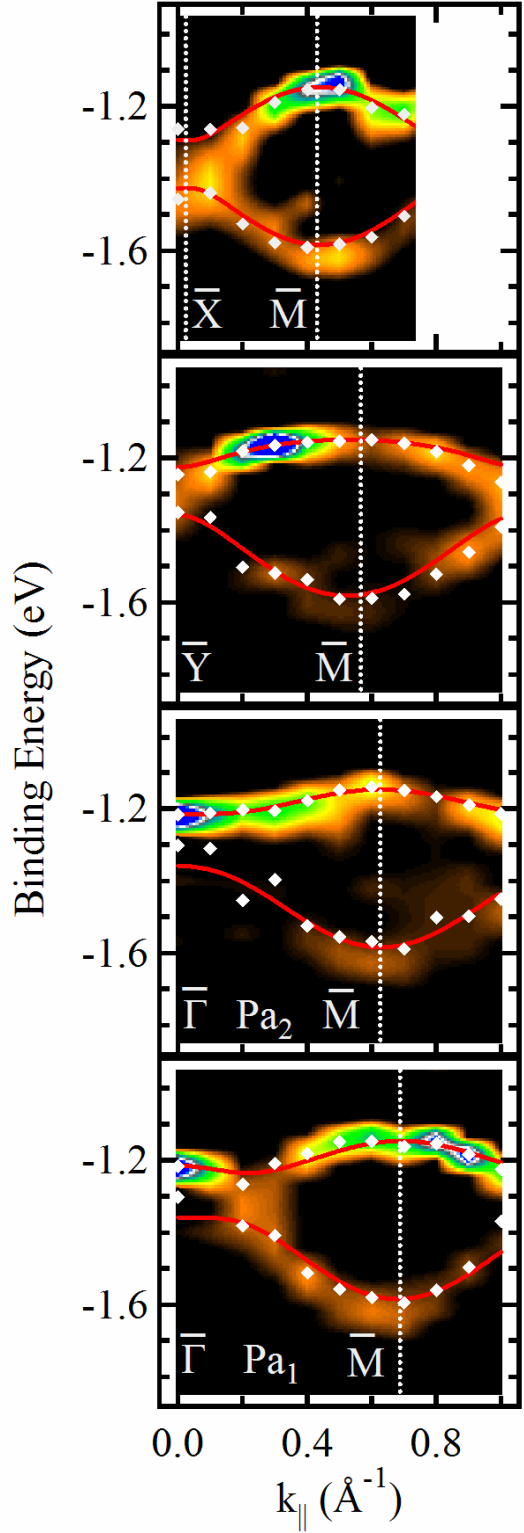

Figure 1 is a compilation of our data along different directions in k-space. The red lines are the result of a simultaneous tight-binding fit to all data points along every direction measured. The error bars of our data are about the size of the data points.

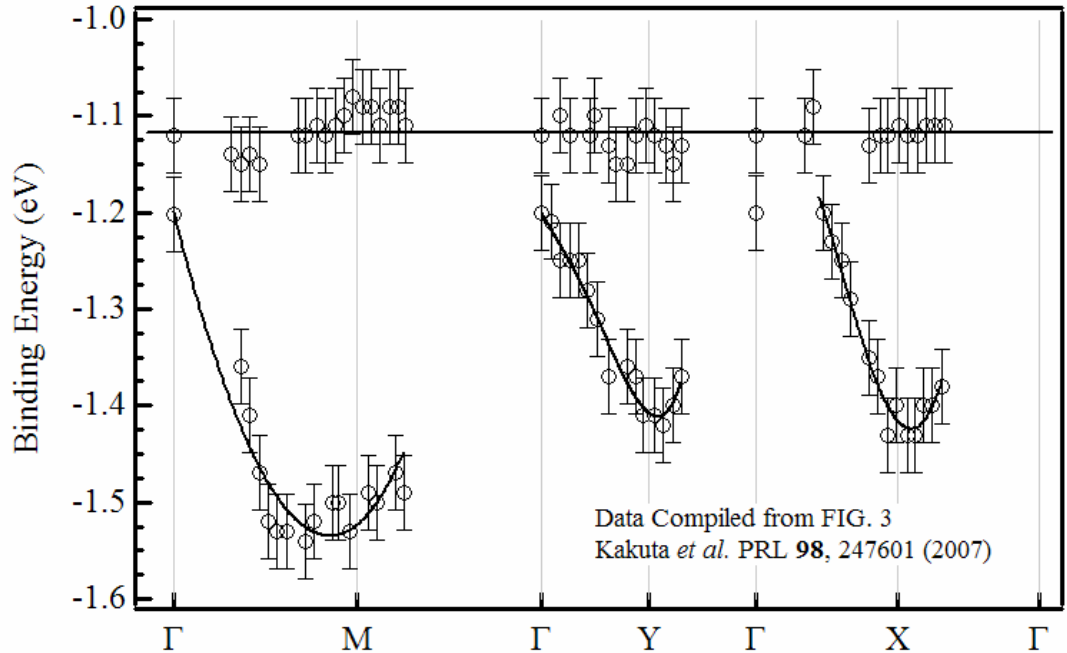

Figure 2 shows data that we extracted from Kakuta's paper. As one can see the E(**k**) peak positions of the HOMO1 band do not show any perceptible dispersion. A fit to these data results in a straight line that intercepts all of their data points within the error bar limits.
For the HOMO2 band we decided to fit a simple polynomial through the individual sections of the data to determine the general trend and to extract the band minima near the $\overline{M}, \overline{Y}$ and $\overline{X}$ points, rather than fitting separate cosine-functions which poorly resemble the data at least in the $\overline{\Gamma}$-$\overline{M}$ and $\overline{\Gamma}$-$\overline{X}$ directions as one can see from Fig. 3(b) in Kakuta's paper. Even though our polynomial fits do not exactly represent the proper dispersion around the $\overline{\Gamma}$ point where one would expect E(**k**) to approach $\overline{\Gamma}$ horizontally, the individual sections are still very useful for the purpose of extracting the band width (difference between maxima and minima) at various high symmetry points.

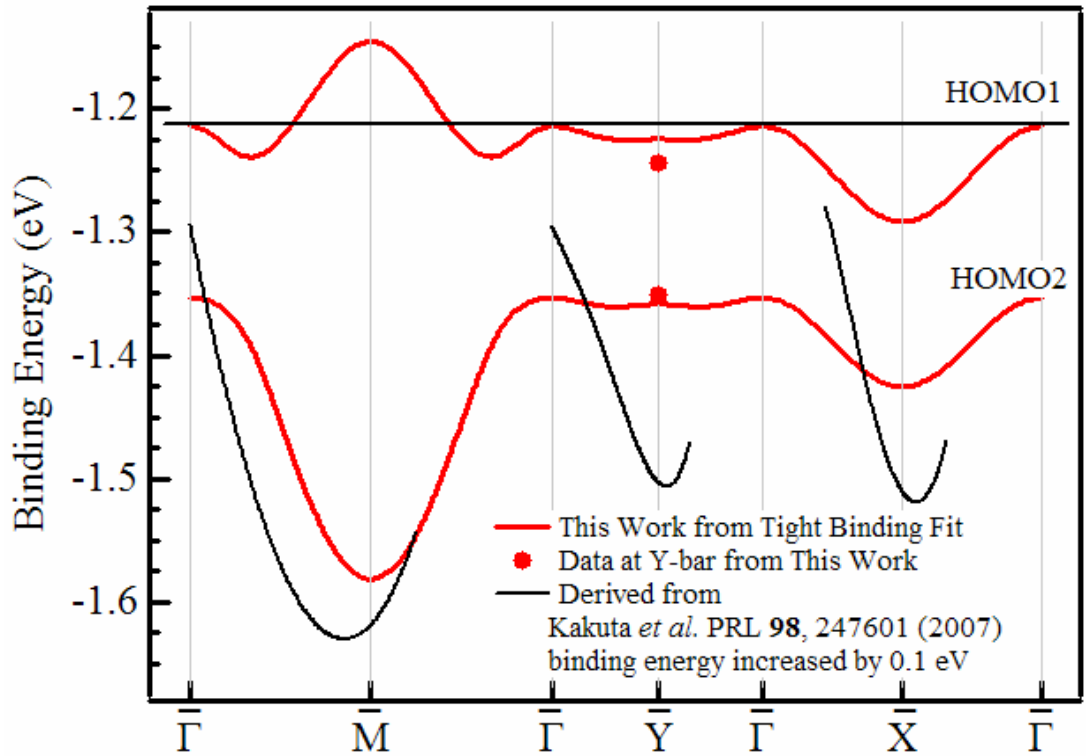

Figure 3 compares the dispersions of the two HOMO-derived bands from our data (red) versus sections of the band structure extracted from Kakuta's paper (black). This figure unambiguously shows that our ARPES analysis allows us to predict the electronic structure of Pn over the entire Brillouin zone to a level of detail and accuracy that is unmatched by previous studies. Also shown as red dots are our data at $\bar{Y}$. Within error bars the data at $\bar{Y}$ are identical to those at $\bar{\Gamma}$ and we therefore found no reason to take additional data along this direction. The apparent non-dispersive nature of the band along this path is in strong contrast to the data reported by Kakuta *et al*.